\documentclass{article}

\usepackage{arxiv}

\usepackage[utf8]{inputenc} % allow utf-8 input
\usepackage[T1]{fontenc}    % use 8-bit T1 fonts
\usepackage{hyperref}       % hyperlinks
\usepackage{url}            % simple URL typesetting
\usepackage{booktabs}       % professional-quality tables
\usepackage{amsfonts}       % blackboard math symbols
\usepackage{nicefrac}       % compact symbols for 1/2, etc.
\usepackage{microtype}      % microtypography
\usepackage{lipsum}		% Can be removed after putting your text content
\usepackage{graphicx}
\usepackage{doi}
\title{Two-photon real-time device for single-particle holographic tracking (Red Shot)}

\author{
Florian Semmer$^{(1)}$, Marie-Charlotte Emperauger$^{(1)}$, Colin Lopez$^{(1)}$, Christine Bogicevic$^{(2)}$,  \\
\And Fran\c cois Treussart$^{(1)}$, Karen Perronet$^{(1)}$ and Fran\c cois Marquier$^{(1)}$
\thanks{francois.marquier@ens-paris-saclay.fr}\\
}

\begin{document}

% \homepage{http:...} %% author's URL, if desired
%%%%%%%%%%%%%%%%%%% abstract %%%%%%%%%%%%%%%%
%% [use \begin{abstract*}...\end{abstract*} if exempt from copyright]

\maketitle

\begin{center}
$^{(1)}$Universit\'e Paris-Saclay, \'Ecole Normale Sup\'erieure Paris-Saclay,
CNRS-UMR9024,
CentraleSup\'elec, LuMIn, \\
Gif-sur-Yvette 91190, France\\

$^{(2)}$Universit\'e Paris-Saclay, CentraleSup\'elec, CNRS-UMR8580, SPMS Lab., 91190 Gif-sur-Yvette, France\\
\end{center}

\begin{abstract}
Three-dimension real-time tracking of single emitters is an emerging tool for assessment of biological behavior as intraneuronal transport, for which spatiotemporal resolution is crucial to understand the microscopic interactions between molecular motors. We report the use of second harmonic signal from nonlinear nanoparticles to localize them in a super-localization regime, down to 15 nm precision, and at high refreshing rates, up to 1.1 kHz, allowing us to track the particles in real-time. Holograms dynamically displayed on a digital micro-mirror device are used to steer the excitation laser focus in 3D around the particle on a specific pattern. The particle position is inferred from the collected intensities using a maximum likelihood approach. The holograms are also used to compensate for optical aberrations of the optical system. We report tracking of particles moving faster than 30 $\mu$m.s$^{-1}$ with an uncertainty on the localization around 40 nm. We have been able to track freely moving particles over tens of micrometers, and directional intracellular transport in neurites.
\end{abstract}

%%%%%%%%%%%%%%%%%%%%%%%%%%  body  %%%%%%%%%%%%%%%%%%%%%%%%%%
\section{Introduction}
Single particle tracking has allowed important advances in the understanding of biological domains such as the fast dynamics of biomolecules at cell membrane\cite{Lee2019}, viral infection\cite{Liu2020} or intraneuronal transport\cite{Kaplan2018}. Most of the experiments consist in extracting trajectories from movies after their acquisition by video-microscopy of biological samples using fluorescent emitters \cite{Godin2017,Haziza2017} or scattering nanoparticles\cite{Kaplan2018,Malkinson2020}. Particles? positions are inferred from each frame of the movies with a few 10s nm localization precision, depending on the number of detected photons\cite{Deschout2014}. The timescale is then given by the frame rate of the movie, from 20 to 100 Hz typically. To achieve such high spatio-temporal resolution, most of the studies are limited to tracking in a single plane of observation.
Video microscopy can be extended to the third dimension (along the direction of propagation $z$), as in interferometric scattering microscopy (iSCAT) \cite{Hsieh2018} or in Point Spread Function (PSF) engineering\cite{Opatovski2021} systems. Both those methods can be based on wide-field microscopy allowing parallel tracking of several particles in 3D and the second one allows even multi-color tracking but they operate over an axial range of only a few 10s $\mu$m. 
Another ensemble of tracking technologies consists in inferring the distance of the emitter to a specific excitation pattern. The recent minimal-photon-fluxes method (MINFLUX) presents the smallest nanometer-range localization precision $\sigma$ that can image single fluorophores in 3D at $\sigma<1$ nm  and track single fluorophores in 2D at a typical 30 Hz  rate, up to 8 kHz for $\sigma\approx 20$ nm \cite{Schmidt2021,Wolff2022}, however the accessible volumes are currently small, less than 500 nm in the $z$-direction, and $\approx 1 \mu m$ in the $x,y$-plane, moreover the method remains very demanding in terms of experimental setup. To our knowledge, none of these methods is reported to be used in optically thick samples. 
To follow the particle over a larger spatial range, other feedback-based 3D real-time single-particle (3D-RT-SPT) tracking methods have been developed \cite{VanHeerden2022}. Mainly, the spatiotemporal-modulation of the excitation light allows to infer the position of the emitter from the analysis of the resulting collected intensity. A feedback loop makes sure that the emitter stays in a excitation area at the microscale, allowing a localization precision in the 10 nm range. Measuring 2D trajectories on long distances and fast can then be done using a 2D excitation pattern that can be executed very fast using Acousto-Optics Deflector\cite{Hou2020} (AOD) or resonant galvo-mirrors\cite{Wehnekamp2019,Annibale2015}. These methods aim to achieve millisecond time-resolution, and a 5-10 nm localization precision but one of the bottlenecks is to be fast also along the optical axis to scan a 3D thick sample and perform precise 3D-localization. In order to pave the way to thick tissues, RT-SPT has been extended to two-photon microscopy, displaying a nice localization precision of about 15 nm at time scales on the order of 20 ms\cite{Levi2005} and more recently at 8 ms\cite{Annibale2015}. A clever multiplexed illumination allowed to lower the time-response below 5 ms in a two-photon system but the maximum tracked speed is finally limited by the response time of a piezostage\cite{Perillo2015}. Being able to realize a 3D tracking in optically thick tissues at nanometer and millisecond dynamics is still difficult and needs new technologies and strategies. %Current techniques use tunable lenses (mainly Tunable Acoustic Gradient\cite{Hou2020}  or electro-tunable lenses\cite{Annibale2015}) to perform millisecond 3D-RT-SPT. %Among those methods, 3D orbital tracking has actually already been used to study mitochondrial transport in living zebrafish larva\cite{Wehnekamp2019}. 

In this paper, we introduce an original 3D-RT-SPT method based on the use of a digital micromirror device (DMD) and Lee holography\cite{Bryngdahl1976,Geng2017} in a two-photon microscope setup. The use of a DMD allows a fast and precise tracking over an extended range. The DMD is used to (i) create a 3D excitation pattern, allowing the super-localization of a non-linear nanoparticle ($\mathrm{BaTiO_{3}}$ nanospheres, $\approx100$ nm diameter), (ii) track the nanoparticle during its motion, changing the 3D excitation-pattern global position, and (iii) correct the wavefront to obtain a diffraction-limited spot at the laser focus. We demonstrate a localization precision lower than 20 nm in the $x,y$-plane, 40 nm in the $z$-direction, and we collected position datas up to 1 kHz acquisition rate. We show that our setup is able to efficiently track moving particles whose speed is as large as ~30 $\mu$m.s$^{-1}$. We finally demonstrate the ability of our setup to track particles internalized in living cells (neuroblasts Neuro-2A) over $\approx$ 20 $\mu$m, displaying highly directional trajectories mediated by molecular motors and characterized by fast and slow phases. Coupling two-photon microscopy, holography and real-time single particle tracking, this work paves the way to deep-tissues measurements of molecular-motors dynamics, kinesin or dynein, whose steps have typical size 10 nm at typical frequency of 100 Hz in physiological conditions along hundreds of $\mu$m, and eventually of intraneuronal-transport parameters\cite{Haziza2017,Kaplan2018,Wehnekamp2019,Peng2022,Wolff2022}.

\section{Results}
\subsection{Experimental setup}
The experiment has been designed to track nanoparticles in biological tissues taking advantage of Second-Harmonic Generation (SHG). This two-photon scattering process allows the conversion of an infrared-light excitation, which is known to be less scattered or absorbed in biological tissues \cite{Golovynskyi2018}, to visible light, for which efficient and sensitive detectors as single-photon counting modules are commonly used nowadays. Such optical probes are expected to be useful in deep tissues. 

\begin{figure}[h!]
\centering\includegraphics[width=15cm]{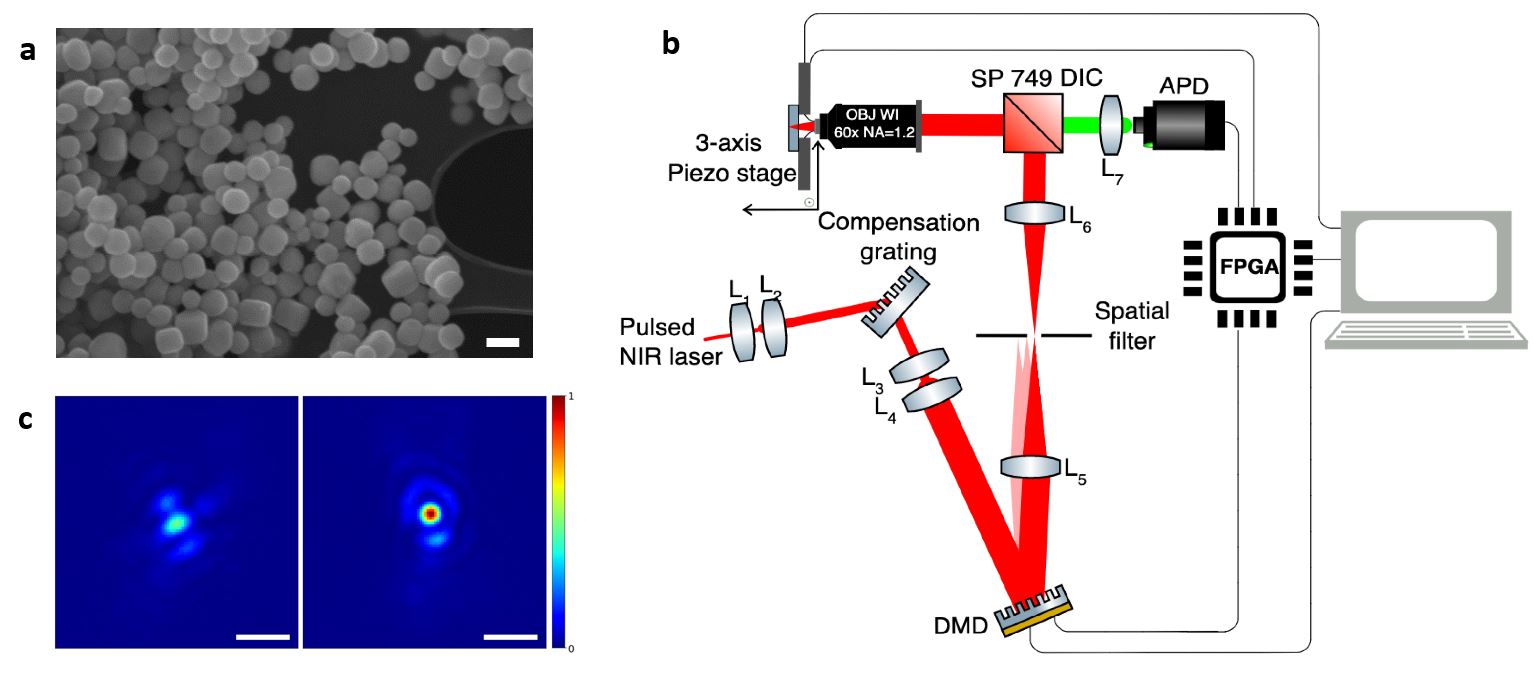}
%\captionsetup{width=1\linewidth}
\caption{DMD and nonlinear nanocrystal tracking setup. (a) SEM image of $\mathrm{BaTiO_{3}}$ nanocrystals used to demonstrate the 3D-tracking. Scale bar 200 nm. (b) Schematics of the DMD-based single particle 3D-tracking setup. A near-infrared (NIR) pulsed laser is first slightly expanded (with $L_{1}$ and $L_{2}$ afocal combination) and then directed towards the DMD after having been pre-compensated for chromatic dispersion with a grating, and  enlarged with a telescope ($L_{3}$, $L_{4}$). After diffraction by the DMD, the -1 order is selected with a diaphragm, recollimated by a lens ($L_{6}$), and sent to the illumination path of a commercial microscope, the yellow dashed line represent the backfocal plane of the microscope objective which is conjugated to the DMD. The beam is then reflected by a shortpass dichroic beamsplitter into a high numerical aperture microscope objective which focuses the excitation laser in the sample. The latter can be translated with a 3-axis piezoelectric stage to extend the tracking area beyond the field of view accessible with the DMD steering. The SHG signal of the $\mathrm{BaTiO_{3}}$ nanocrystal is collected through the same objective, transmitted through the BS and focused ($L_{7}$) on the active area of a single-photon counting module (SPCM). The experiment is interfaced by custom LabVIEW FPGA and LabVIEW programs.  (c) Two different beam profile before and after phase correction. Profile imaged after reflection on a mirror in the focal plane of the objective on camera. The intensities are normalized to the maximum intensity of the corrected profile. Scale bar 3 $\mu$m.}
\label{fig:1}
\end{figure}

We used $\mathrm{BaTiO_{3}}$ nanoparticles (NP), diameter $\approx 100$ nm (Fig. \ref{fig:1}a) having a large nonlinear susceptibility which leads to a good SHG efficiency (typical two-photon scattering cross section $2.10^{3}$  GM for a 100 nm diameter NP \cite{Kim2013}). Figure \ref{fig:1}b shows a schematic diagram of the setup. It relies on a conventional Lee holography configuration adapted to a two-photon microscope, inspired by a paper from Geng et al.\cite{Geng2017}. We use an infrared pulsed laser (titanium-doped sapphire laser Ti:Sa, Chameleon Ultra-II; Coherent, USA) having a 80 MHz repetition rate and a pulse width of 140 fs. We tuned its emission wavelength at $\lambda=1028$ nm to excite SHG from $\mathrm{BaTiO_{3}}$ NP at $\lambda/2=514$ nm. The excitation laser beam is directed to a DMD (Superspeed V-7001 ; Vialux, Germany) before being sent to an inverted microscope (Eclipse Ti-E; Nikon, Japan) and focused into the sample using a water immersion high numerical aperture microscope objective corrected for a 170 $\mu$m glass coverslip (Plan Apo VC 60xA/1.20 WI; Nikon, Japan). 

As we use a spectrally broad laser, we need to compensate the DMD chromatic dispersion to avoid an enlargement of the spot at the focus of the microscope objective. We realize a conjugated dispersion with a compensation grating (T-1000-1040, 1000 lines/mm ; LightSmyth, USA) and two lenses $L_{3}$ and $L_{4}$. These lenses form an afocal system which optically conjugates the compensation grating with the DMD, and obtain the desired angular magnification to optimally compensate the chromatic dispersion \cite{Geng2017}. A first telescope (lenses $L_{1}$ and $L_{2}$) is placed before the compensation grating to fit the beam diameter to the DMD active area. Each micromirror on the DMD can be tilted with an angle of $\pm 12^{\circ}$ and the intensity in the desired diffraction order is maximized choosing carefully the angle of incidence on this device to fulfill a blazing criterion. We chose this angle of incidence to be equal to $0^{\circ}$ because it minimizes the sensitivity of the chromatic dispersion to the incident angle on the DMD. The blazed order direction is thus at an angle of $24^{\circ}$ from the incident beam. The beam diffracted by the hologram displayed on the DMD is focused (with $L_{5}$) onto a pinhole to select only the (-1) order of the displayed diffraction pattern. The lens $L_{6}$ associated to $L_{5}$ completes a last afocal system which ensure the conjugation of the DMD with the back focal plane of the microscope objective with the adapted magnification. As a result,  the spot at the focus of the microscope objective is the Fourier transform of the pattern displayed on the DMD.

The sample is coupled to a 3D piezo-stage (LT3-200 ; Piezoconcept, France) to extend the range of tracking beyond the field of view made accessible with the DMD holograms. The SHG from the tracked non-linear nanocrystal is collected by the same objective, then transmitted through a short-pass dichroic beamsplitter (FF749-SDI01 ; Semrock, USA) and focused ($L_{7}$) onto a single photon counting module (SPCM-AQRH-13; Excelitas, USA). All the apparatuses are interfaced with a FPGA board (PCIe-7820; National Instrument, USA) and the experiment is controlled by a home-made program written in LabVIEW and LabVIEW FPGA.

\subsection{Aberration correction}
The DMD presents some flatness imperfections that are at the origin of the main laser-beam wave-front deformations and lead to a non-ideal Point Spread Function (PSF). These defects can be observed by placing a mirror in the objective focal plane and conjugating the surface of the mirror to a camera sensor (fixed at one side port of the microscope): the focused laser spot, resulting from the residual excitation light transmitted by the dichroic BS, displays strong astigmatism. Such defects must be corrected before using the device for single-particle tracking. To this aim, we take advantage of digital holography to apply common tools of adaptive optics. We adjust the coefficients of the Zernicke polynomials (mainly the coefficients associated to vertical and horizontal astigmatism) to maximize the peak intensity of the gaussian and obtain ultimately a diffraction-limited PSF. Figure \ref{fig:1}c displays the PSF before and after correction of astigmatism.Quantitatively, the corrected PSF has a Strehl ratio of around 0.8, meaning that the standard deviation of the incident wave front is less than $\lambda/14$ compared to a flat wave front \cite{vandenBos00}. In comparison the Strehl ratio obtained without correction is around 0.2. When corrected, the The final excitation intensity can be up to 600 $\mathrm{kW.cm^{-2}}$.
Because we perform 3D tracking, we finally used a single nonlinear nanocrystal to probe the two-photon PSF in 3D via its SHG response, by scanning the crystal position in the $(x,y,z)$ directions with the piezostage (see supplementary figure 1).

\subsection{Holography and positioning}
By changing the spatial frequency of a periodic pattern imprinted on the DMD we can position the laser spot at different locations in the focal plane of the objective. It is also possible to encode a spherical phase in the hologram and thus modify the focusing position along the optical axis\cite{Geng2017}. The focused spot $S$ of the laser beam can thus be placed at different locations $\mathbf{r}_S =(x_{S},y_{S},z_{S})$ within the sample along the three dimensions. S can be seen as the image of a virtual source point C located at $\mathbf{r}_C =(x_C,y_C,z_C)$ through the whole optical system. The hologram to be imprinted onto the DMD is the binarized pattern resulting from the interference between the spherical wave that would be emitted by C and a reference laser light (plane wave whose wave vector is denoted $\mathbf{k}_{inc})$. Such a hologram can be computed at each pixel position $\mathbf{r}(x,y)$ of the DMD surface as: 
\begin{equation}
   H(x,y)=\left\lfloor \frac{1}{2} \left[1+\cos\frac{2\pi}{\lambda}\left(\frac{\mathbf{k}_{inc}.\mathbf{r}}{\|\mathbf{k}_{inc}\|}+\| \mathbf{r}-\mathbf{r}_{C}\|-\varphi(\mathbf{r})\right)\right]\right\rfloor
\end{equation}

Where $\lfloor X\rfloor$  is the integer part of $X$ and $\varphi(\mathbf{r})$ the phase map that takes into account aberration corrections. In addition, care must be taken  to ensure there is no overlap between the useful diffraction order of the hologram and a natural diffraction order of the DMD grid. It leads to an off-axis configuration which can be interpreted as the choice of a carrier frequency in Lee holography to define the center of the positions attainable by the excitation laser.  %To avoid noise, we want the center of the holographic field of view to be relatively far from a diffraction order of the DMD grid. We are in off-axis holography configuration. This apply a general grid to the hologram with a specific spatial frequency called carrier frequency in Lee holography. Modification of this frequency leads in a modification of the focalization position in the image space of the objective and local phase shift of this carrier grid code for a phase which can be a defocus and an aberration compensation phase map.

\subsection{Localization precision}
\begin{figure}[h!]
\centering\includegraphics[width=13cm]{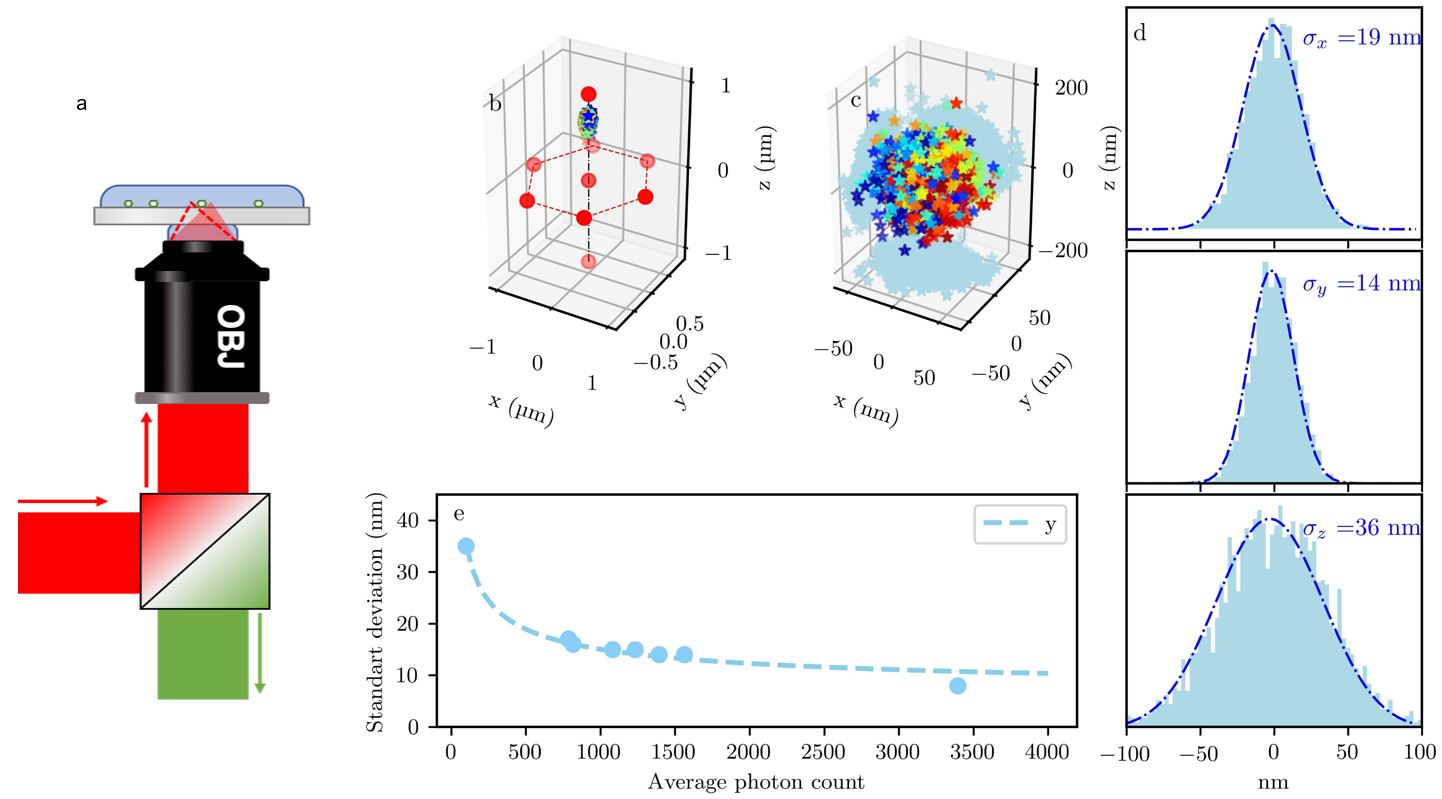}
%\captionsetup{width=1\linewidth}
\caption{Measurement of the localization precision of an immobile single non-linear nanocrystal. (a) Configuration of the inverted microscope: NPs (green dots) deposited on a glass coverslip are illuminated by an infrared laser whose focus point varies in time (two configurations represented in dotted lines and in plain fill, The SHG is collected through the objective to a SPCM). (b) The excitation pattern is represented by the red dots, (c) The zoomed scatter plot shows the dispersion of the estimated position of an immobile nanocrystal: each color is a different measurement, for this set of data, each point corresponds to an illumination of 4.5 ms. (d) Histograms showing the distribution of the three position coordinates. The point-dotted line is a gaussian fit whose standard deviation $\sigma$ is given in each subplot, (e) standard deviation of the y position measurement as a function of the collected photon number N during a set of excitations (sum of the photon numbers collected for the 9 excitation positions). The dotted line decreases as $\frac{1}{\sqrt{N}}$.}
\label{fig:2}
\end{figure}

The position of the NP is inferred from a set of excitation spots positioned around and in close proximity of it. We first address the precise localization of a fixed emitter. We use a fixed $\mathrm{BaTiO_{3}}$ NP deposited on a coverglass and covered with water to be in the best working conditions of the microscope objective (Fig. \ref{fig:2}a). We sequentially display different holograms on the DMD, each of them focusing the beam at a known position in the vicinity of the NP (Fig. \ref{fig:2}b). In the SHG process, the emitted number of photons depends quadratically on the intensity of the excitation laser, hence related to the distance between the excitation point and the position of the NP. Using a maximum likelihood approach \cite{Balzarotti2017} and the knowledge of the 3D-PSF of the setup, we are able to precisely estimate the location of the emitter in the excitation pattern. The latter consists of 9 spots, distributed as follows: 6 spots are regularly placed on a circle in a plane orthogonal to the optical axis, describing a hexagon, and 3 spots are along the optical axis, located on a segment crossing the hexagon at its center as shown in figure \ref{fig:2}b. With numerical simulations discussed in the next paragraph, the radius of the circle encompassing the hexagon has been optimized to be $a =$ 470 nm \cite{Wang2010}, and the distance between two consecutive points on $z$ is $\mathrm{d}z =$ 1 $\mu m$.

We set first the photon detection integration time at 500 $\mu$s at each position of the pattern, so that the total integration time for one localization of the observed NP is 4.5 ms leading to a typical number of $10^3$ collected photons ($\approx 2.2\times 10^5$~photons.s$^{-1}$). We determine the localization precision with repetitive position measurements, as displayed in figure \ref{fig:2}b and c. Its value is characterized by the standard deviation of the set locations, yielding 19 nm and 14 nm along the $x$ and $y$ directions respectively and 36 nm in the optical axis $z$ direction (Fig. \ref{fig:2}d). Increasing the integration time leads to larger photon numbers and thus a better localization precision. Figure \ref{fig:2}e shows the evolution of the position-measurement standard deviation along the y-axis as a function of the total photon number N received from the set of 9 excitation points. We observe the expected $1/\sqrt{N}$ behavior above an asymptotic value of around 10 nm, that we attribute to remaining vibrations of our setup, that could be corrected in the future.

Note that from a temporal point of view, the DMD has a maximum refresh rate of 22 kHz corresponding to a minimum response time of 45 $\mu$s. The overall rate results on one hand from the dwell time defined as the sum of the refreshing time and the integration time of the detector (during which the same hologram must be displayed), and on the other hand from the number of holograms needed to infer the position of the NP (9 in this work). As usual, for a targeted localization precision the photon budget is the limiting factor\cite{Deschout2014}, hence fixing the maximum measurement rate. In the following, we tracked NP whose SHG maximum emission was between $10^{5}$ photons.s$^{-1}$ and $5\times 10^{5}$  photons.s$^{-1}$. %The laser spot position can thus be changed on that timescale within the field of view regardless of the starting point or the traveling distance. This specificity is called inertia free positioning and  is also made possible with other devices, in particular Acousto-Optics Modulators used to perform random access calcium imaging\cite{Akemann2022}\cite{Nadella2014}.  

\subsection{Tracking strategy}

As we cannot perform the maximum likelihood estimation in real time we evaluate the position of the emitter with a centroid computation program embarked on the FPGA board. However, when the NP move too far from the center of the excitation pattern, the localization uncertainty increases and either the emitter needs to be moved toward the center of the pattern or the pattern itself can follow the particle. Most of the existing RT-SPT techniques adopt the first strategy and act on the relative position of the emitter with respect to the pattern at every step of the tracking algorithm using a piezo stage\cite{Wehnekamp2019,Balzarotti2017,Hou2020,Perillo2015}. We chose the second strategy and change the excitation pattern accordingly to the NP position evolution, as it provides a faster scan speed, compared to the piezo stage whose response time is of a few tens of milliseconds (in close-loop). In order to use the DMD at its full speed we preloaded a library of holograms associated to positions regularly spaced in a 3D cubic volume (see Supp. Mat. \ref{ExcPatt})]. The number of excitation positions is limited by the RAM of the DMD control board to 87380 holograms.

A convenient and compact predefined mesh is a hexagonal mesh grid, as the one used for the precision of localization. Considering the parameters $a$ and d$z$ previously defined previously, and easy-to-use memory management, a mesh of $21\times 21 \times 5$ holograms has been chosen,  setting the boundaries of a holographic tracking volume of around $9.9 \times 8.5\times 5$ $\mu$m$^3$ which limits then the total field of view. When the NP reaches the boundaries of this volume, we use the piezo stage to re-center the NP in the tracking volume. It is finally possible to track the NP over longer distances when it is needed. During the piezo motion, we do not measure the position of the emitter. The tracking algorithm consists finally in the simplified following steps: (i) Localize a NP, (ii) when the NP reaches a given limit into the excitation pattern, use the closest next pattern, (iii) move the piezo-stage if the limit of the field of view is reached. The maximum likelihood estimation of the position at each step is performed afterwards from the recorded intensities to get the best precision from our setup. %Note that a specific faster algorithm has also been developed to run a modified maximum likelihood estimator on FPGA board to find the NP in real time\cite{Balzarotti2017}.  We did not adapt this real-time estimator to our setup yet as our localization sequence is based on more excitation positions which makes it computationally heavier and we could not make it fit in the FPGA memory.

%\section{Characterizations}
\subsection{Controlled trajectories}
\begin{figure}[h!]
\centering\includegraphics[width=12cm]{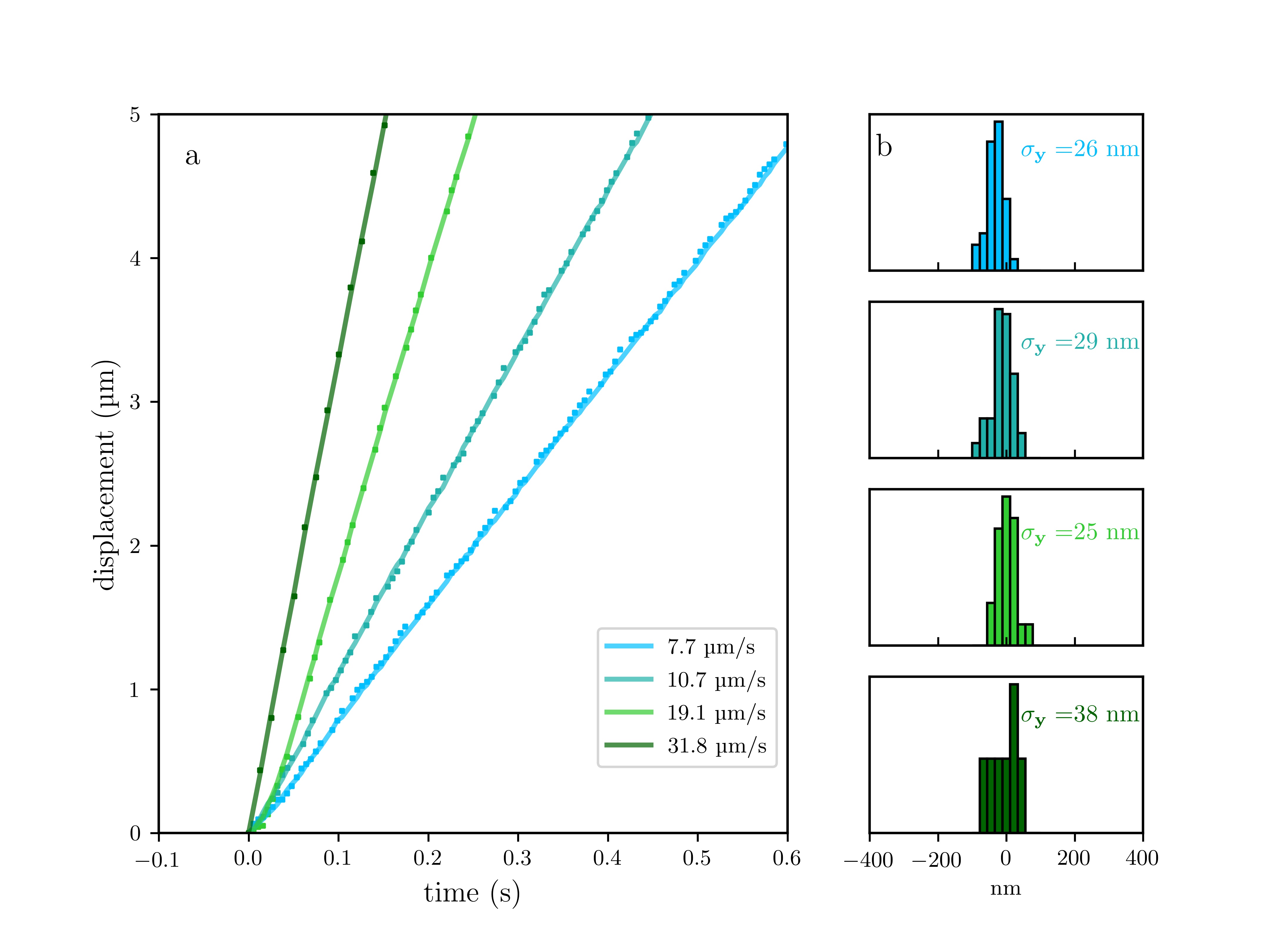}
%\captionsetup{width=1\linewidth}
\caption{(a) Tracks of a nanocrystal of $\mathrm{BaTiO_3}$ moved by a piezostage at different velocities, inside the holographic field of view but through several pattern changes. (b) Histograms of the error to the piezo stage position. Standard deviation of the distribution for velocities of 7.7, 10.7, 19.1 and 31.8 $\mu$m.s$^{-1}$ are respectively 26, 29, 25 and 38 nm. The localization precision remains almost constant for velocities up to $\approx 20 \mu$m.s$^{-1}$. }
\label{fig:3}
\end{figure}

We fix the integration time at 500 $\mu$s at each excitation point (220 Hz localization rate). To illustrate the spatiotemporal resolution of the setup, we first tracked a NP fixed on a coverglass and moved by the piezo stage along a known predefined trajectory. The tracked NP emits an average of $1.5\times 10^5$~photons.s$^{-1}$. Figure \ref{fig:3}a shows examples of tracks of NP moving along the $y$-axis with different velocities. We observe that the position measurement match the expected true trajectory of the piezostage represented in plain lines.
The distributions of errors is shown on the histograms displayed in figure \ref{fig:3}b. Interestingly, the mean error does not strongly depend on the velocity in the range 7 to 31 $\mu$m.s$^{-1}$. The typical error is most of the time less than 30 nm, in very good agreement with the expected value from localization precision measurements of Fig. \ref{fig:2}d, taking into account the difference of photons number emission. For a 100 $\mu$s integration time per excitation point (1.1 kHz localization rate), the typical error stays relatively low at 40 nm for the same NP (see Supp. Mat. \ref{fastacq}). As expected, decreasing the integration time  lowers the photon budget and degrades the localization precision so that brighter NP in a two-photon process\cite{Liu2020,Peng2022} would definitely help using the full speed of the DMD.

\subsection{Natural trajectories}
\subsubsection{Nanoparticles in solution}
\begin{figure}[h!]
%\centering\includegraphics[width=15cm]{Figure_Diffusion_resolue.jpg}
\includegraphics[width=13cm]{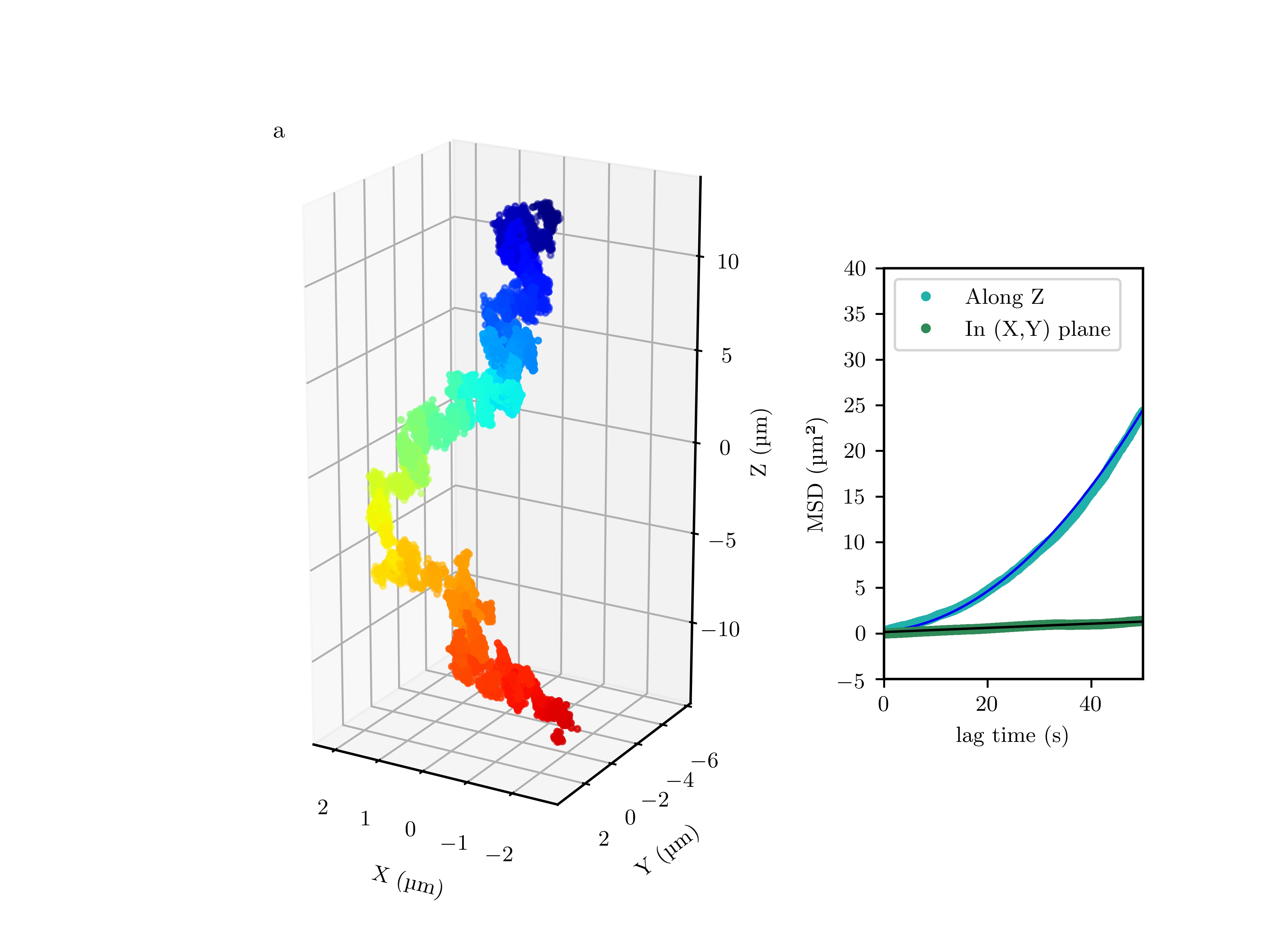}
%\captionsetup{width=1\linewidth}
\caption{(a) 3D Trajectory of a freely moving particle in a water/Glycerine mixture at a localization rate of around 220 Hz. The colormap corresponds to the time, from blue at the beginning of the trajectory to red at the end of the trajectory. (b) Mean Square Displacement (MSD) in the $x,y$ plane, and in the $z$ direction. The former is fitted with a linear model, whereas the latter is fitted with a parabolic model (solid lines). The linear behaviour is characteristic of a purely diffusive motion.}
\label{fig:4}
\end{figure}
To check the setup ability to track a NP over a large field of view, we use freely moving NP in a water/glycerine mixture. Tracks can be acquired over tens of micrometer on several minutes. .
Figure~\ref{fig:4}a shows a typical trajectory of a given NP in 3D during ~4 min. We observe a clearly diffusing behaviour in the $x,y$ plane, whereas the motion is more directional along the $z$ direction. We first focus on the trajectory in the $x,y$ plane only. We derive the Mean Square Displacement (MSD) from the data sets $x_i=x(t_i)$ and $y_i=y(t_i)$ at discrete times $t_i$ as: 
$$
\mathrm{MSD}\left(\Delta t_{ij}=t_j-t_i \right)= \langle x_j^2-x_i^2+y_j^2-y_i^2 \rangle -\left\langle\sqrt{x_j^2-x_i^2+y_j^2-y_i^2 }\:\right\rangle^2
$$
Where the braket stands for an average over the same lag times for a given trajectory.

Figure \ref{fig:4}b displays the MSD as a function of the lag time $\Delta t$ for the typical trajectory depicted in Fig. \ref{fig:4}a. As expected from a diffusive motion, we observe a linear behavior whose slope is related to the diffusion coefficient $D$: $\mathrm{MSD}(\Delta t)=4D\Delta t$. We are thus able to extract a diffusion coefficient from the measurement. We estimate $D\approx 0.011 \mu\mathrm{m^2.s^{-1}}$ which is consistent with the NP size and viscosity range considering the Stokes-Einstein relation for a spherical particle $D=k_B T/(6? \eta R_{\mathrm{NP}})$ where $k_B$ is the Boltzmann constant, $\eta$ the viscosity and $R_{\mathrm{NP}}$ is the NP radius. 

Along the $z$ direction, the behavior of the NP is more complex to interpret as the motion becomes directional: the NP goes upwards in the liquid, keeping a random Brownian motion. In that case the MSD can be written as $\mathrm{MSD}(\Delta t)=2D \Delta t+v^2 \Delta t^2$, where $D$ is the diffusion coefficient previously measured in the $x,y$ plane and $v$ is the mean velocity of the directional motion. Simulations show that this behavior is compatible with an effect of the so-called scattering optical force from the excitation laser (see Supp. Mat. \ref{Simus}). We measure a velocity of $\approx$94 nm.s$^{-1}$ that can be related to an average optical force pushing the NP of $F=6\pi \eta R_{\mathrm{NP}}=k_B T v/D$, whose order of magnitude is then $\approx 3.4\times 10^{-2}$~pN, much higher than the weight of the NP, around 0.2 fN. If such a force perturbs the free motion in the fluid, the order of magnitude is negligible compared to the force that a molecular motor could apply to an endosome embedding such a NP, around 10 pN, so that we believe our tracking technique is fully available for measuring directional transport in cells. 

\subsubsection{Trajectories in living cells}
The tracking method has finally been tested on NP internalized in living cells displaying directional trajectories and typical go and stop phases. We used mouse neuroblasts (Neuro-2A) cells 2D cultures and NP were added to the cultured medium of the cell (see Supp. Mat. \ref{N2Acult}). 
\begin{figure}[h!]
\centering\includegraphics[width=15cm]{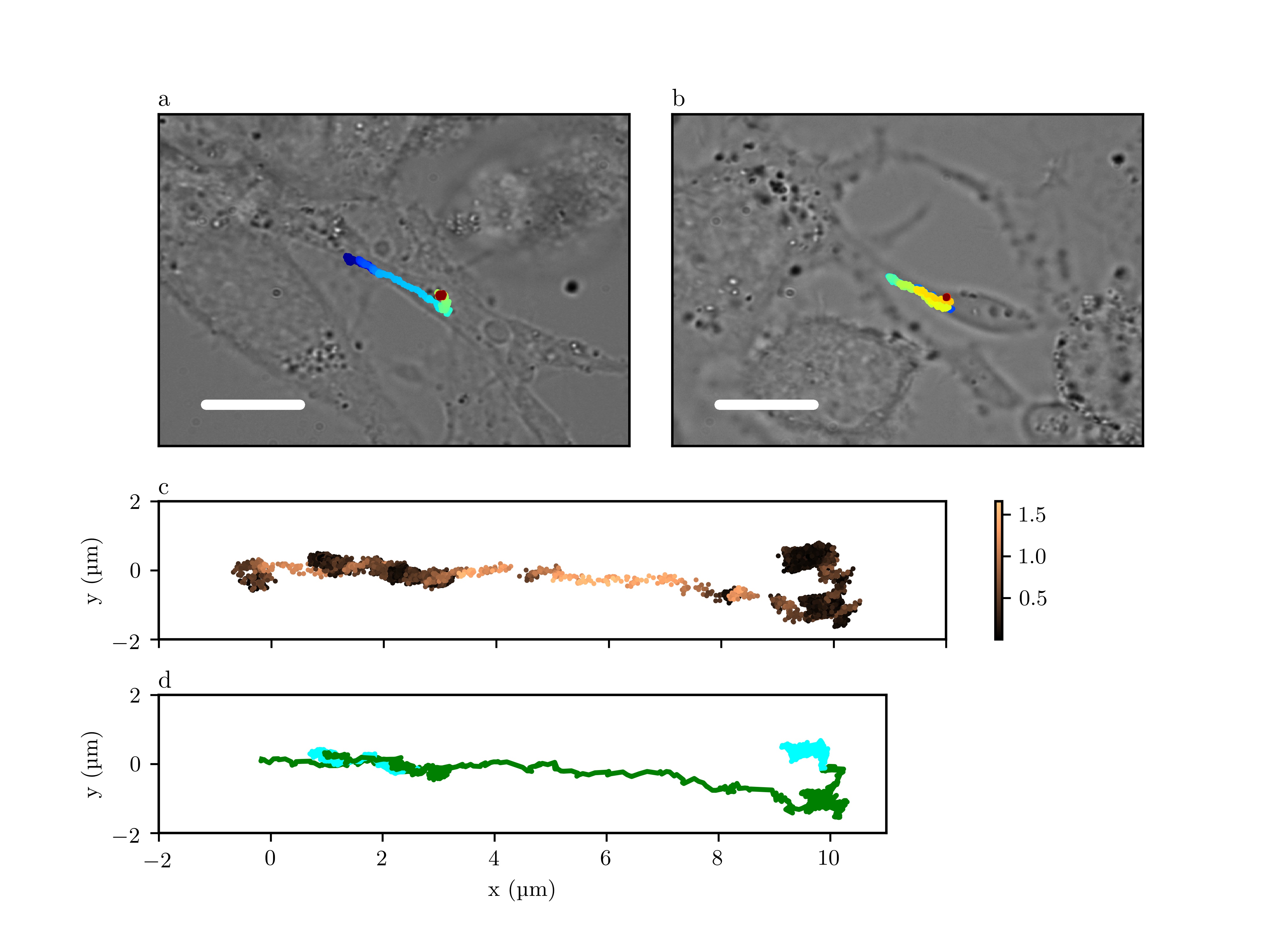}
%\captionsetup{width=1\linewidth}
\caption{(a and b) Two trajectories observed in living Neuro-2A cells during 2 min. The colormap corresponds to the time from blue at the beginning of the trajectory to red at the end of the trajectory. Trajectories are superimposed with microscopy images of the corresponding cells. The white line is 10 $\mu m$. (c) Part of the trajectory from (a), each position of the NP is associated to a color corresponding to its instantaneous velocity. The colorbar is in  $\mu$m.s$^{-1}$. (d) Part of the trajectory from (a) the colors stand for the back and forth movements. }
\label{fig:N2A}
\end{figure}
After a few minutes, some NP are likely to be spontaneously internalized in cells by endocytosis\cite{Haziza2017}. This is confirmed by the trajectories observed for the NP. Figures \ref{fig:N2A}a and \ref{fig:N2A}b display two highly directional trajectories, acquired during 2 min, superimposed with microscopy images. The directional behavior cannot be attributed to a diffusive motion in the medium or on the cell membrane\cite{Chen2016}. In contrast this is the expected motion of an endosome containing a NP and taken over by molecular motors\cite{Cui2007,Haziza2017,Chou2022}. We focus on the latter trajectory on fig. \ref{fig:N2A}c, where the positions of the NP are represented in the $x,y$ plane with a color corresponding to its instantaneous velocity. 

We now clearly see slow and fast phases usually associated to stop and go states of the dynamics of endosomes. The velocities order of magnitude, around 1 $\mu$m.s$^{-1}$, and up to 2 $\mu$m.s$^{-1}$ are also in the expected range of such processes\cite{Kural2005}. Depending on the molecular-motors family (kinesin or dynein) predominantly involved in the transport process, we can also observe some back and forth movements (Fig. \ref{fig:N2A}d). During the experiment, no alteration of the cells has been observed. We hence believe that this setup could be used to track NPs in living cells for intraneuronal transport measurements. Due to the two-photon capability, it has the potential to be adapted to the observation of NP transport in more complex tissues as for instance in zebrafish larvae\cite{Malkinson2020} in a superlocalization regime and at a very high rate. 

\section{Conclusion}
In conclusion, we have presented a new two-photon 3D Real-time Single particle tracking method based on digital holography mediated by a DMD. We demonstrated the ability of our setup to localize fixed nanoparticles with a precision of less than 20 nm in $x$ and $y$ directions and 40 nm along the $z$ direction depending on the number of collected SHG photons. We have shown that we can acquire trajectories with a time resolution down to 1 ms and a typical localization precision of 30 nm along $x$ and $y$ directions and 60 nm along $z$ direction. coupled to a piezostage, we tracked free nanoparticles in solution over 10s of micrometer along all directions, test our tracking device on biological sample (living neuroblasts Neuro-2A) and observed typical directional trajectories driven by molecular motors. 
Aiming to apply the tracking in thick samples we currently work on an adaptive optics loop to compensate for aberration induced by the sample itself. 

\section{Supplementary materials}

\subsection{3D PSF}
\begin{figure}[h!]
\centering\includegraphics[width=7cm]{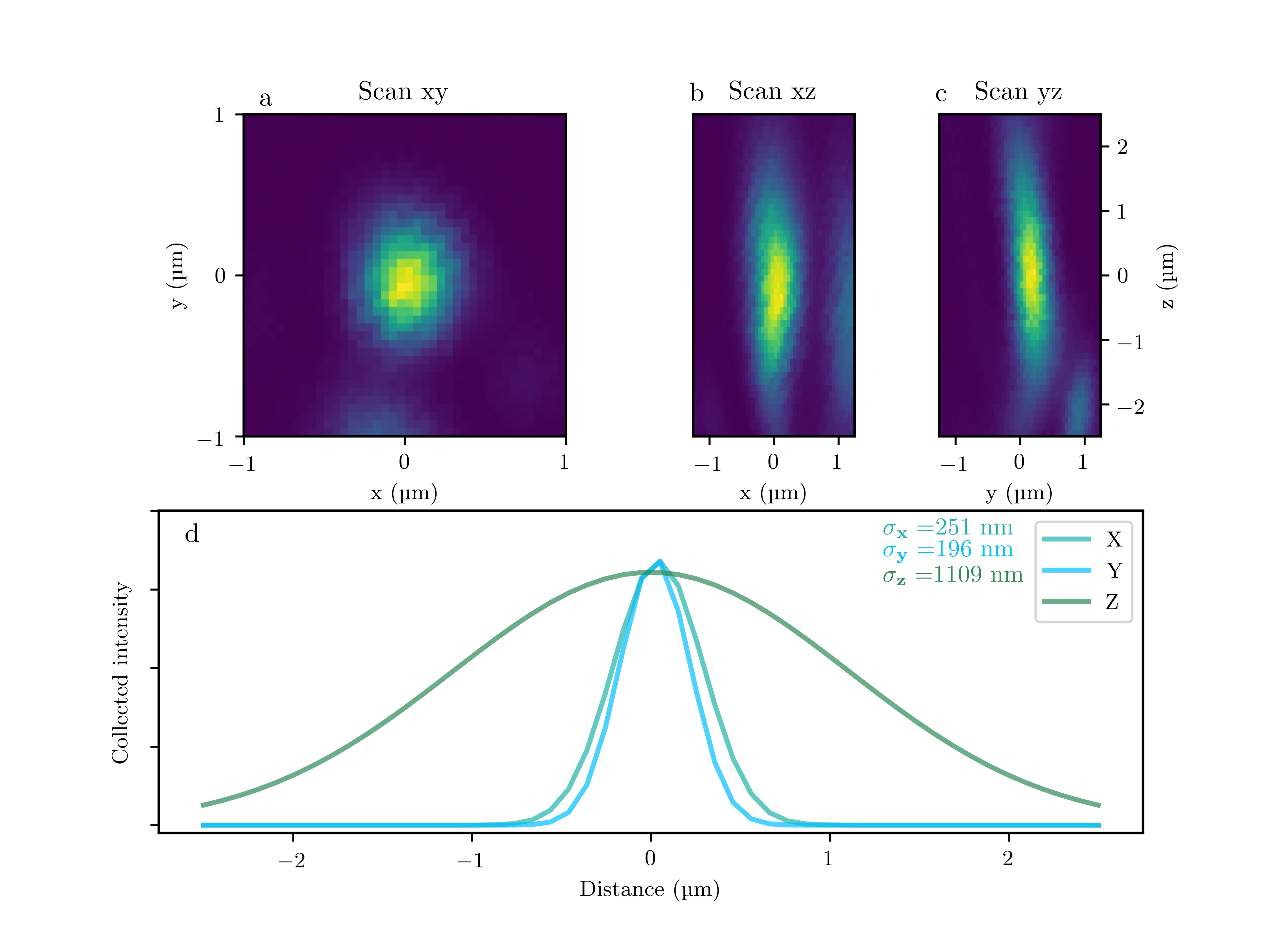}
%\captionsetup{width=1\linewidth}
\caption{a)b)c) 2D Scan of a single $\mathrm{BaTiO_{3}}$ fixed on a coverslip, colors stand for the collected SHG intensity. d) Gaussian fit of sections of the previous scans,the standard deviations are respectively 251,196,1109 nm.}
\label{fig:Fig Sup PSF 3D}
\end{figure}
The knowledge of the 3D- detection point spread function (PSF) of the setup is crucial to use a maximum likelihood approach \cite{Balzarotti2017} and to be able to precisely estimate the location of the emitter in the excitation pattern.
In order to access this information we scan a single $\mathrm{BaTiO_{3}}$ nanoparticle fixed on a coverslip by moving it with the piezo stage and collecting the SHG signal. Fig \ref{fig:Fig Sup PSF 3D} shows three 2D scans of the same particle and sections of theses scans adjusted with Gaussian function. In this case we observe a little ellipticity in the XY plane with $\sigma_{x}=251$ nm and $\sigma_{y}=196 $nm, this difference in the XY can be explain by the shape of this nanoparticle. Along the z direction the standart deviation is 1109 nm which is close to what is expected in \cite{Zipfel2003} if we consider the detection point spread function (PSF) and the radius at $1/e^{2}$ .

\subsection{Excitation pattern}\label{ExcPatt}
To use the DMD at its full speed we can only display holograms that has already been loaded into the RAM of the DMD controller. This is one of the drawbacks of the use of a DMD because if one wants to acquire fast, it can not ask for a continuous repositioning of the excitation pattern. Hence we need to think about the arrangement of all the possible location we want to focus the laser at. We compute holograms library to form an hexagonal mesh in 3D and we decide to change the holographic excitation pattern, made of holograms from the library, when the particle is $0.6a$ away from the center or $0.6\mathrm{d}z$ away from the plane of the hexagon. The question of an optimal solution for $a$ and $\mathrm{d}z$ seem natural. In comparison, for techniques such as orbital tracking where the excitation spot move continuously along a circular trajectory, the optimum for the parameter $a$ is found to be the value which defines a circular excitation area with uniform power and is $w_{0}$/$\sqrt{2}$.  In our case, What we want is to minimize the localization bias during Real-time measurement when the particle it at the boundary to avoid instabilities. We simulate localisation at the boundary for different values of $a$ and $\mathrm{d}z$ with different background level and found that $a$ = 470 nm and $\mathrm{d}z$ = 1000 nm seemed to be an optimum in our case. 

\subsection{Scattering optical force from the excitation laser}\label{Simus}
To be updated

\subsection{Controlled trajectories }\label{fastacq}
\subsubsection{Fast acquisition}

\begin{figure}[h!]
\centering\includegraphics[width=7cm]{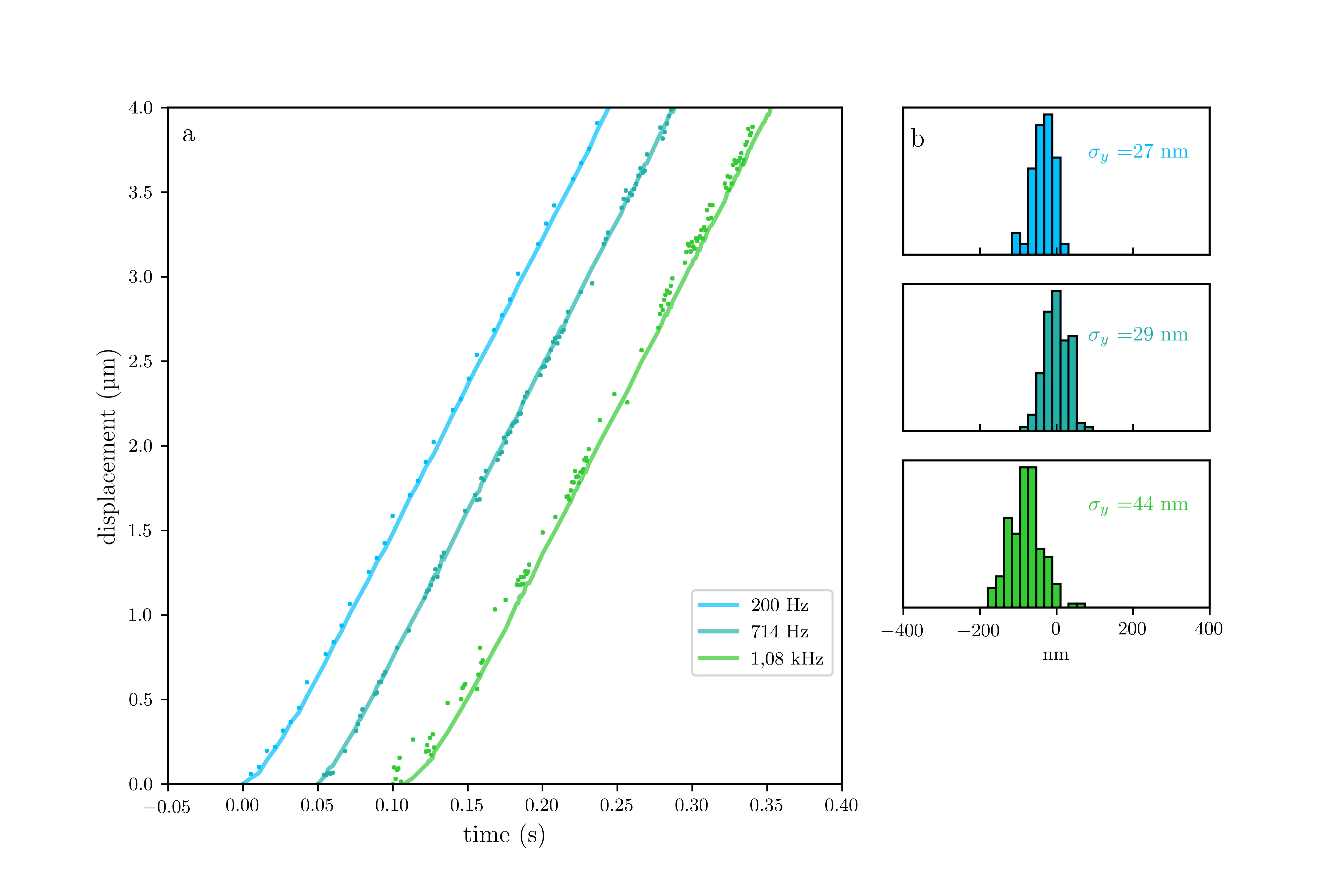}
%\captionsetup{width=1\linewidth}
\caption{a) Three trajectories of the same $\mathrm{BaTiO_{3}}$ nanoparticle at 17 $\mu m.s^{-1}$ moved by the pizeo stage along the x direction and with different acquisition rate : 200Hz, 714Hz and 1.08kHz.b) Histograms of the error to the piezo stage position. Standard deviation of the distribution are respectively  27, 29 and 44 nm. }
\label{fig:Sup x Freq}
\end{figure}

Several acquisitions were made increasing the acquisition rate. In Fig\ref{fig:Sup x Freq} we tested values of 200Hz, 714Hz and 1,08kHz by choosing integration times of respectively 5 ms, 1.4 ms and 0.9 ms. We observe that the error to the piezo stage position is still around 30 nm until 714Hz, for faster acquisition it raises 44nm. Of course the number of collected photons lower when the frequency rises. We point out that the sequence update takes 8.5 ms approximatively so when we acquire much faster and if the np moves fast, we see that the trajectory is not uniformly measured.
Depending on the trajectory expected in biological medium we can adapt the parameters to find a compromise between high spatial resolution and fast acquisition.

\subsubsection{$x$ and $z$ trajectories}

\begin{figure}[h!]
\centering\includegraphics[width=7cm]{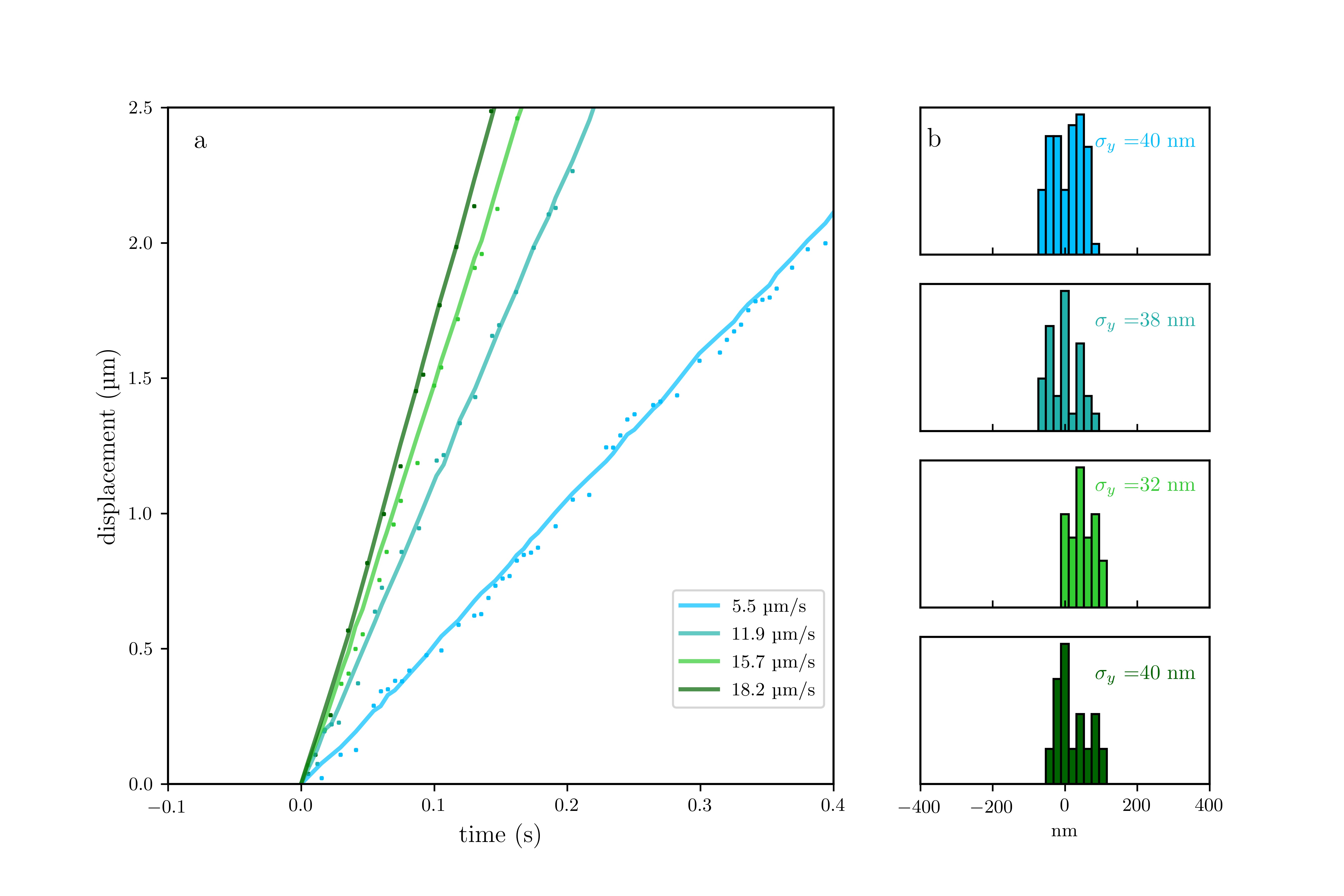}
%\captionsetup{width=1\linewidth}
\caption{a) Tracks of a nanocrystal of $\mathrm{BaTiO_{3}}$ moved by a piezo stage along the x direction at different velocities with 200Hz acquisition time. b) Histograms of the error to the piezo stage position. Standard deviation of the distribution are respectively 40,38,32,40 nm.}
\label{fig:Sup x}
\end{figure}

\begin{figure}[h!]
\centering\includegraphics[width=7cm]{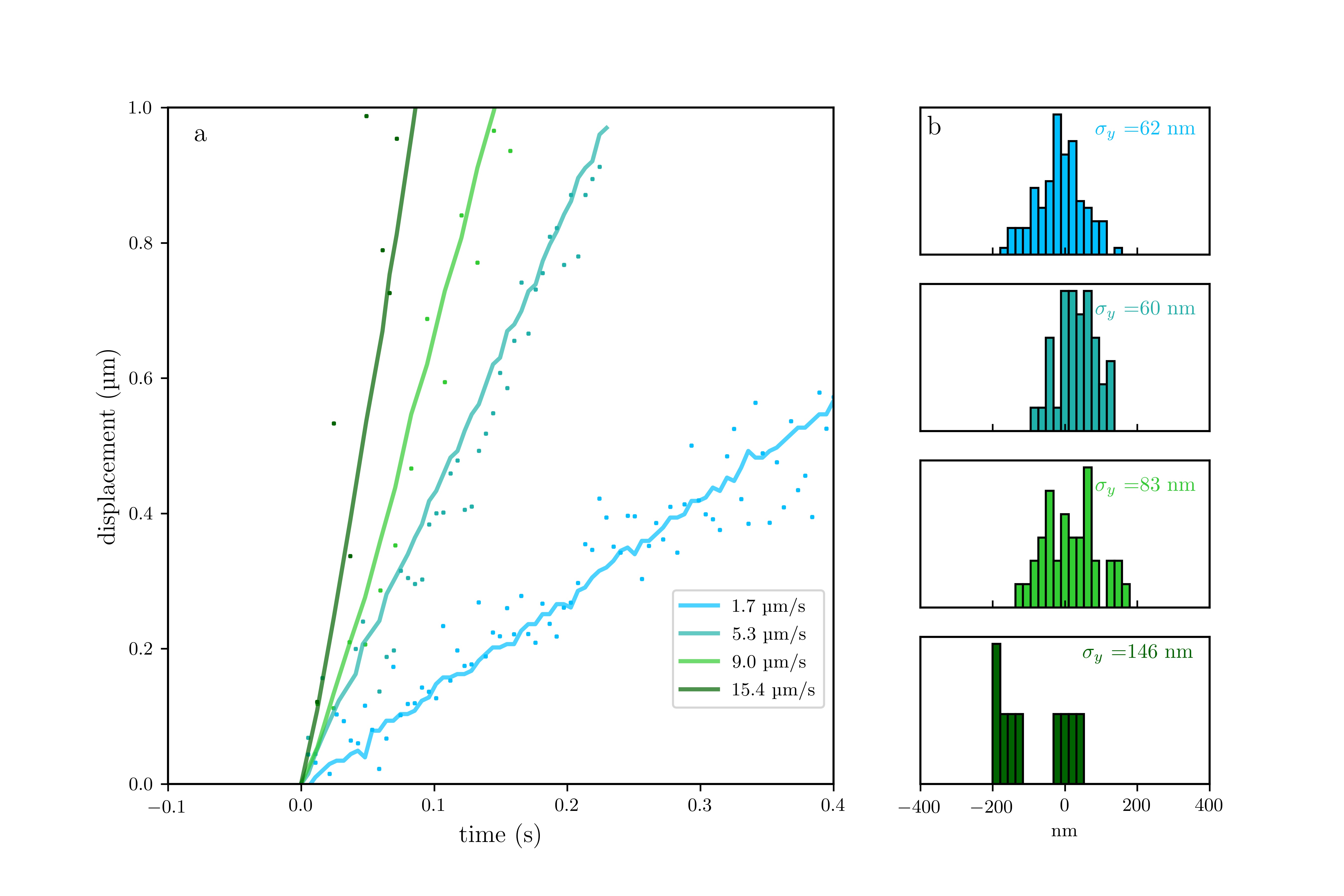}
%\captionsetup{width=1\linewidth}
\caption{a) Tracks of a nanocrystal of $\mathrm{BaTiO_{3}}$ moved by a piezo stage along the z direction at different velocities with 200Hz acquisition time. b) Histograms of the error to the piezo stage position. Standard deviation of the distribution are respectively 62,60,83,146 nm.}
\label{fig:Sup z}
\end{figure}

\newpage
\subsection{N2A culture}\label{N2Acult}

Neuro-2A cells are mouse neuroblasts with neuronal and amoeboid stem cell morphology isolated from brain tissue. They were divised and deposit on a glass coverslip of 18mm in a culture well with cultured medium (DMEM 10$\%$ Fetal Bovine Serum and 10$\%$ Pen-Strep antibiotic) on day one. Four days after, the solution of spherical BTO 38 nanoparticules at $300mg/mL$ was placed into ultrasonic bath during 30 min and added to the cultured medium of the cell. Each well received 20$\mu L$ of solution and was stored again into incubator for one night. The glass coverslip was deposit on the sample carrier on the $5^{th}$ day and was covered by 400$\mu L$ of new culture medium.

\section{Backmatter}

\textbf{Funding}
: This work has received financial support from the French National Research Agency ANR (ANR-18-CE09-0027), and by a public grant overseen by the ANR as part of the ?Investissements d?Avenir? program (reference: ANR-10-LABX-0035, Labex NanoSaclay). The research was also supported by a ENS Paris-Saclay scholarship.\\

\textbf{Acknowledgments}
: We thank Christelle Langevin, Guillaume Baffou and Patrick Chaumet for fruitful discussions. We thank Jean-Pierre Mothet, Martina Papa and Baptiste Grimaud for her help in the N2A culture cell. We thank Simon L'Horset and Sebastien Rousselot for their help in electronics and mechanics. \\

\textbf{Disclosures}
: The authors declare no conflicts of interest.

\section{References}

%%%%%%%%%%%%%%%%%%%%%%% References %%%%%%%%%%%%%%%%%%%%%%%%%

%%%%%%%%%% If using BibTeX:
\bibliographystyle{alpha}
\bibliography{biblio.bib}
%\printbibliography
\end{document}